\documentclass[12pt]{article}                   

\usepackage{amsmath}    
\usepackage{amsfonts}   
\usepackage{amssymb}
\usepackage[dvips]{graphicx}   
\def\pa{\partial}
\def\<{\langle}
\def \>{\rangle}

\def\rp{\rho_{3}}
\def\rl{\rho_{1}}

\def\f{\frac}
\def\Rs{R_{\odot}}
\def\parc{\f{3\gamma-4}{\gamma-1}}
\def\pard{\f{\gamma}{\gamma-1}}
                  
\title{On Solar Radius  Variation with magnetic Activity } 
\author{Arnab Rai Choudhuri \\ 
and \\ 
Piyali Chatterjee \\
Department of Physics, Indian Institute of Science, Bangalore-560012, India. \\
({email:arnab@physics.iisc.ernet.in}, {piyali@physics.iisc.ernet.in})}
\date{\today}
\begin{document}          
\maketitle
\begin{abstract}
In response to the claim by Dziembowski {\em et al.}\ (2001) that the
solar radius decreases with magnetic activity at the rate of 1.5
km yr$^{-1}$, we consider the theoretical question whether a radius
variation is expected with the solar cycle. If the radius variation
is caused by the magnetic pressure of toroidal flux tubes at the
bottom of the convection zone, then the dynamo model of Nandy and
Choudhuri (2002) would suggest a radius decrease with magnetic
activity, in contrast to other dynamo models which would suggest
a radius increase.  However, the radius decrease is estimated to be
only of the order of hundreds of metres.
\end{abstract}

\section{Introduction}
By analyzing the f-mode frequency data from MDI for the period
1996--2001, Dziembowski {\em et al.}\ (2001) claimed 
that the Sun shrinks in radius from the solar minimum to the solar
maximum, at the rate of about 1.5 km per year.  If correct, then this
will certainly be a most dramatic discovery in solar physics.  
However, this claim has not yet been supported by other researchers, who
often present conflicting results.  Antia (2003) provides a review of
the subject and concludes that there is no compelling evidence
for solar radius change.  It may be noted that some earlier studies
(N\"oel, 1997; Emilio {\em et al.},\ 2000) reported much larger radius
variation (up to 700 km) based on other (non-helioseismic)
methods.  It is clear that more work is needed
before a definitive conclusion can be reached.
The aim of this paper is to discuss whether on theoretical grounds
we expect a solar radius variation with solar cycle.  Other than
some early work by Spruit (1982, 1994), we have not been able to find any
other theoretical analysis of this problem.  In the light of recent
developments of solar physics, this theoretical question certainly
needs revisiting.

There are two ways in which a magnetic field present
in the solar interior may lead to an increase of solar radius.
\begin{enumerate}
\item The pressure of the magnetic field embedded in the gas may 
cause the gas to expand, thereby inflating the size of the Sun.
\item The inhibition of convection by the magnetic field may cause
a piling up of heat below the magnetic field.  This would result
in heating of the surrounding regions, leading to an expansion.
\end{enumerate}
From both of these arguments, it appears that the radius of the
Sun should increase when more magnetic fields are present.  Since
we normally think of the solar maximum as being the time when the
Sun has more magnetic fields in the interior, we would expect the
Sun to be larger in size during the solar maximum.  Shrinking of
the Sun with increasing magnetic activity, therefore, would seem
improbable at the first sight. 

We shall argue that the recent solar dynamo model of Nandy and
Choudhuri (2002), in principle, allows for the possibility
of shrinking of the Sun during the solar maximum.  The important
theoretical question is whether the theoretically expected radius
decrease agrees with the claim of Dziembowski {\em et al.}\ (2001). Spruit 
(1982, 1994) studied the effect of heat blocking by magnetic fields
and concluded that the total radius variation cannot be more than
0.14 km. We mainly look at the other effect of magnetic pressure
and present order-of-magnitude estimates, suggesting that the
radius variation, for reasonable assumptions, cannot be as large as
what Dziembowski {\em et al.}\ (2001).  We have also made a mathematical
formulation of the problem, to go beyond order-of-magnitude estimates.
To our surprise, we found that the mathematical problem is much 
more difficult to solve than what we anticipated.  Although our
mathematical analysis so far has not yielded very definite results,
we present a brief account of it to help future researchers who may
wish to analyze this problem in greater depth.

Section 2 addresses the question whether we expect the radius
to increase or decrease during the solar maximum.  An order-of-magnitude
estimate is presented in Section~3, showing that a much smaller
variation of radius is expected than what is reported by Dziembowski
{\em et al.} \ (2001). Our incomplete mathematical analysis is given in
Section~4.  Finally, our conclusions are summarized in Section~5.

\section{Increase or decrease of radius with magnetic activity?}    

Both the effects listed above---expansion due to magnetic pressure
and heat blocking by inhibition of convection---would lead to an
increase of the solar radius when there is more magnetic flux.
So it appears that the Sun can shrink with increasing magnetic 
activity only if there is less magnetic flux inside the Sun at
the time of solar maximum.  We know that the Sun has much more magnetic
flux near the surface in the form of sunspots during the solar 
maximum.  It used to be tacitly assumed that the magnetic flux
in the interior of the Sun should also have a peak during the
solar maximum.  Some recent dynamo models, however, raise doubts 
about this assumption.

It has been realized for some time that the meridional circulation
plays an important role in the solar dynamo (Choudhuri {\em et al.}, \ 1995).
This meridional circulation is poleward near the surface and must be
equatorward at the bottom of the solar convection zone
(SCZ).  With the mapping
of angular velocity distribution in the solar interior by helioseismology,
it is clear that the strongest gradient of angular velocity occurs at
high latitudes within the tachocline and that is the most likely region
for the creation of toroidal field.  To explain why active regions
appear at low latitudes, Nandy and Choudhuri (2002) have developed
a solar dynamo model in which the meridional flow penetrates a little
bit below the tachocline and advects the strong toroidal field (created
at high latitudes inside the tachocline) to low latitudes through 
convectively stable layers below the tachocline. Only when the
meridional circulation rises in the low latitudes, the toroidal fields
are brought within the SCZ and rise quickly due to magnetic
buoyancy to form sunspots.  In the model of Nandy and Choudhuri (2002),
therefore, the toroidal flux which erupts at the solar surface was
created a few years earlier at high latitudes within the tachocline,
since it took time for the flux to be transported from there.  If the
meridional flow speed at the base of the SCZ is taken to
be of order 1.2 m s$^{-1}$ (Hathaway {\em et al.}, 2003), then the time taken
for flux transport should be in the range 5-10 yr.   Hence the flux
should have been created a few years before the solar maximum. In this
scenario, the solar maximum is
the time when the Sun merely gets rid of the flux stored below
the SCZ.  In contrast to the traditional dynamo models in
which during the solar maximum the Sun generates the strong 
toroidal field and the interior flux has a peak value, the model of
Nandy and Choudhuri (2002) suggests the solar maximum to be the time
when the magnetic flux in the Sun is reduced after being created a few
years earlier.  Although the solar maximum is the period when the magnetic
flux near the surface in the form of sunspots
peaks, it may also be a time when the magnetic
flux at the bottom of the SCZ {\em actually decreases}.
So, to settle the question whether a magnetic effect would lead to
increase of decrease of the solar radius, we need to ascertain whether
the magnetic effect is caused by the magnetic flux near the surface
or the magnetic flux at the bottom of the SCZ. We point out
that several simulations of active region formation
(Choudhuri and Gilman, 1987; Choudhuri, 1989; D'Silva and Choudhuri, 1993;
Fan {\em et al.}, \ 1993; D'Silva and Howard, 1993; Caligari {\em et al.}, \ 1995) have
pinned down the value of the magnetic field at the bottom of
SCZ to be around $10^5$ G. 

We first make a few remarks on Spruit's analysis of the heat blocking
problem (Spruit, 1982, 1994).
If the magnetic field is sufficiently strong to suppress convection,
the effects of heat blocking become independent of the magnetic field
(i.e.\ it does not matter if the magnetic is just strong enough to
suppress the convection or much stronger than that).  On the other
hand, results of radius variation due to magnetic pressure depend
critically on the value of the magnetic field, as we shall see below.
Spruit (1982) mainly considered heat blocking by vertical flux tubes
near the solar surface.  Although we believe in 
the existence of strong toroidal flux
tubes with $10^5$ G magnetic field at the bottom of 
SCZ, heat blocking by these flux
tubes will not have much effect for two reasons.  Firstly, due to the
high concentration of magnetic field inside these flux tubes, 
the flux tubes would occupy a smaller volume compared to the volumes of
flux tubes below the sunspots.  Hence convection would be suppressed
in much smaller volumes.  Secondly, the bottom of the SCZ is the
region where convective heat transport just begins to win over the 
radiative transport.  A suppression of convection there is not expected
to have as much effect as it would have in the interior of SCZ.  We
thus conclude that we can still trust Spruit's results and 
his conclusion that the radius variation can at most be of
order $2 \times 10^{-7} \Rs$, which is only $0.14$ km. It is
not only considerably
less than what is reported by Dziembowski {\em et al.} \ (2001), it also
has the wrong sign.  More heat blocking below sunspots takes place
during the solar maximum and the radius should increase at that time.

If magnetic flux decreases at the bottom of SCZ during the solar
maximum, as suggested by the model of Nandy and Choudhuri (2002),
then the magnetic pressure is reduced and it is possible that the
Sun shrinks due to this at the time of the solar maximum.  We now
turn to making an estimate of how much this radius decrease is 
expected to be.
   
\section{Basic considerations and rough estimates}

We now consider radius variations due to magnetic pressure of
the concentrated flux tubes at the bottom of SCZ. When the magnetic 
flux is created, it disturbs the hydrostatic equilibrium. It is 
well known that the time scale for restoration of hydrostatic equilibrium
is very short (see, for example, Schwarzschild, 1958, p.\ 32).  As
Schwarzschild (1958) comments: ``There is no doubt that punishment
would follow swiftly for any disobedience of the hydrostatic law.''
We, therefore, assume that the Sun would fairly quickly adjust itself
to a new hydrostatic equilibrium when the magnetic field is created.
So we shall basically compare hydrostatic equilibria with and without
magnetic fields embedded inside, without bothering about the
adjustment time scales.  Our approach thus is going to be quite
different from the approach of Spruit (1982) while studying the heat
blocking problem, in which two time scales (the diffusion time and the
Kelvin-Helmholtz time) play important roles.

To calculate the expansion of a flux tube due to magnetic pressure,
we need to specify the thermodynamic conditions. If the temperature
in the interior of the flux tube is the same as the outside
temperature, then certainly the flux tube expands due to 
magnetic pressure (see, for example, Choudhuri, 1998, \S14.7).
However, such a flux tube also tends to be buoyant and to rise
against gravity.  The flux tubes are believed to be stored in
the mildly sub-adiabatic region immediately below the bottom of
the SCZ.  If a buoyant flux tube rises in a region of sub-adiabatic
temperature gradient, it can be easily shown that it becomes
less buoyant (Parker, 1979, \S8.8; Moreno-Insertis, 1983). One
expects that the flux tube would rise until it becomes neutrally
buoyant, i.e.\ until the interior temperature decreases sufficiently
to make the density inside and outside exactly equal.  If the
flux tube is neutrally buoyant, then its density is no different
from the outside density and there is no increase in the overall
size of the system.  Thus, if all the magnetic flux inside the
Sun remained neutrally buoyant, then there would be no change in
the solar radius at all.  In view of our lack of understanding of the physics
of flux tubes at the bottom of SCZ, let us allow for the possibility
that this may not be the case.  The penetrating meridional circulation
proposed in the model of Nandy and Choudhuri (2002) would drag
the flux tubes downward, making them hotter and not allowing them
to come up by magnetic buoyancy.  Considering a favourable circumstance
that the flux tubes are not colder than the surroundings, let us see
if we can have sufficient radius increase of the Sun. 

We now make a rough estimate by considering a one-dimensional
atmosphere in rectangular geometry. Suppose
a horizontal magnetic field $B$ is created within this atmosphere
in a layer between $z$ and $z+ L$. Because of the magnetic pressure
inside this layer, the gas pressure would be less than the 
pressure $p$ that would exist there if the magnetic field were not
present. To find the corresponding density decrease, we assume
the region with the magnetic field to have the same temperature
as the surrounding region.  Then we can give the same
arguments which we give for flux tubes in thermal equilibrium with
surroundings (see, for example, Choudhuri, 1998, \S14.7) and conclude
that the density decrease is given by
$$\frac{\delta \rho}{\rho} \approx \frac{B^2}{8 \pi p}. \eqno(1)$$
In a one-dimensional atmosphere,
such a density decrease implies that the magnetic layer would be
inflated by an amount $\delta L$ compared to thickness which the
gas in this layer would have in the absence of the magnetic field.
Hence, when the magnetic field is created, the
overlying atmosphere will be pushed up by this amount
$\delta L$, which is given by
$$\frac{\delta L}{L} \approx \frac{\delta \rho}{\rho}. $$
Note that we are not bothering about the signs of $\delta \rho$
and $\delta L$.
Then from (1),
$$\frac{\delta L}{L} \approx \frac{B^2}{8 \pi p}. \eqno(2)$$
Since
$$F = B L$$
is the magnetic flux through the magnetic layer (per unit length
of the layer measured in a direction perpendicular to $B$ in the
layer), we can write (2) as
$$\delta L \approx \frac{F^2}{8 \pi p L}. \eqno(3)$$
It is clear from (3) that the rise of the atmosphere is more if the
same flux $F$ is concentrated in a narrower layer (i.e.\ if the magnetic
field is stronger). 

Taking $p = 5 \times 10^{13}$ dyn cm$^{-2}$ at the bottom of the SCZ,
(2) tells us that for magnetic field created there we must have
$$\delta L \approx 7.7 \times 10^{-6} \left( \frac{B}{10^5} \right)^2 L.
\eqno(4)$$
For magnetic field of order $10^5$ G, we expect $\delta L/ L$ to be
only of order $10^{-5}$.

We now estimate the radius shrinkage of the Sun due to the release of magnetic
flux from the interior during the solar maximum.  Assume that there are
about $10^4$ eruptions during the maximum and each eruption brings out
flux in the range from $10^{20}$ Mx to $10^{21}$ Mx.  Then the total flux
coming out of the interior of the Sun is about $10^{24}$ -- $10^{25}$ Mx.
This flux comes from the bottom of SCZ, where the magnetic field is
$10^5$ G.  This gives a cross-sectional area of $10^{19}$ -- $10^{20}$
cm$^2$. To compare with the one-dimensional model, we distribute
this flux in a uniform shell at the bottom of SCZ.  Taking $L$ to be
the thickness of this shell,
the cross-sectional area through which the magnetic field passes  
should be equal to $\pi R_{\rm bot} L$, where
$R_{\rm bot}$ is the radial distance of the bottom of SCZ.  Equating this 
cross-sectional area to $10^{19}$ -- $10^{20}$ cm$^2$, we get
$L$ in the range from about $10^8$ cm to $10^9$ cm (i.e. from about 1000 km
to 10,000 km).  It follows from (4) that the expected maximum shrinkage
of the Sun can only be in the range from 0.1 km to 1 km. If this radius shrinkage
takes place over 5 years, then even the most favourable value 1 km of radius
shrinkage will give a shrinkage rate of only 0.2 km per year. According to
(3), the magnetic field at the bottom of SCZ has to be squeezed in a layer
narrower by one order of magnitude if the radius shrinkage rate had to be
as large as what Dziembowski {\em et al.}\ (2001) find.  This would imply the
horizontal magnetic field at the bottom of SCZ should be of order $10^6$ G.
Such a strong field would create problems with the theoretical explanation
of Joy's law (D'Silva and Choudhuri 1993).  

It should be kept in mind that the radius shrinkage rate estimated above
is an over-estimate.  If the magnetic flux tends to become neutrally
buoyant in some regions, then certainly the radius variation will be less.
Additionally, Choudhuri (2003) has argued that magnetic flux would
exist in concentrated form only in limited regions at the bottom of SCZ,
the field being more diffuse in other regions.  Since diffuse magnetic
field would not cause much radius variation, in accordance with (3),
the actual radius variation would be smaller if concentrated magnetic field
regions are intermittent.

\section{A hydrostatic model with a magnetic layer}

After the order-of-magnitude estimate, we now try to construct hydrostatic models 
of the SCZ, with and without magnetic
fields in the interior, to determine the distance by which the outer surface
shifts between these two models.  After considering a hydrostatic model without
magnetic fields, we shall assume that the magnetic field is created in a shell
$r_b < r <r_t$ and calculate the new equilibrium.  To our great surprise, we
found this problem to be much more complicated than what we expected.  Although
we have not been able obtain very definite results, we discuss the formulation
of the problem to provide guidance for future researchers who may wish to tackle
the problem with more realistic models.

\setcounter{equation}{4}
Let us first consider the case without the magnetic field and construct a
hydrostatic model from the bottom of the overshoot layer at  $R_{BCZ} = 0.67 R_{\odot}$
to the surface at $R_{\odot}$. Let $p_0(r), \rho_0(r), 
T_0(r)$ be the pressure, density and temperature structures in this case.
The equation for hydrostatic equilibrium is given by 
\begin{equation}
\label{eq:hydro}
\f{d{p_0(r)}}{d{r}} = -\f{\rho_0(r) GM_{\odot}}{r^{2}}.
\end{equation} 
The temperature gradient is slightly sub-adiabatic in the overshoot layer and
slightly super-adiabatic in the SCZ.  We assume an adiabatic stratification
throughout and use the equation of state
\begin{equation}
\label{eq:statef}
p_0(r) = \kappa {\rho_0(r)}^{\gamma}.
\end{equation}
Substituting (\ref{eq:statef}) in (\ref{eq:hydro}) and solving
the resulting ODE with the boundary condition $\rho_0(R_{\odot})=0$,
we have the following expressions for $\rho_0(r), p_0(r)$, 
and $T_0(r)$. To compute the temperature we have used the ideal gas law
$p=\rho R_{gas} T$.
\begin{eqnarray}
\label{eq:rpt}
\begin{pmatrix}
\rho_0(r) \\ 
p_0(r) \\ 
T_0(r)\\ 
\end{pmatrix}
=
\begin{pmatrix}
[\f{\gamma-1}{\gamma} \f{GM}{\kappa} \{\f{1}{r}- \f{1}{R_{\odot}}\}]
^{\f{1}{\gamma -1}}\\
\kappa[\f{\gamma-1}{\gamma} \f{GM}{\kappa} \{\f{1}{r}- 
\f{1}{R_{\odot}}\}]^{\f{\gamma}{\gamma -1}}\\
\f{\gamma-1}{\gamma}GM\{\f{1}{r}-\f{1}{R_{\odot}}\}\\
\end{pmatrix} 
\end{eqnarray}

We now consider that a magnetic field $B(r)$ is created in the shell $r_b < r <r_t$, i.e.\
the magnetic field remains zero in the layers below and above. 
The hydrostatic equation is satisfied
piecewise in all the three layers.
The subscript '$1$'
indicates quantities below the magnetic shell, the subscript '$2$' denotes quantities
inside the magnetic layer, whereas '$3$' 
indicates those above. In the field-free regions '1' and '3', pressure and density
should satisfy the hydrostatic equation (5) and the adiabatic equation (6)
(with the subscripts '1' and '3' replacing the subscript '0'). Then the solutions
also will be similar to (7).  We can assume the layer below to remain unchanged
after the introduction of the magnetic field.  So $\rho_1(r)$, $p_1(r)$ 
and $T_1(r)$ are equal to $\rho_0(r)$, $p_0(r)$ 
and $T_0(r)$ below $r = r_b$.  In the case of the upper layer, we assume that 
outer radius is now $R$ instead of $\Rs$.  So $\rho_3(r)$, $p_3(r)$ 
and $T_3(r)$ can simply be obtained from (7) by replacing $\Rs$ by $R$
(which essentially means changing the constant of integration), i.e.\
\begin{equation}
\rp(r) = \Big [\f{(\gamma-1)GM_{\odot}}{\gamma \kappa}
\Big (\f{1}{r}-\f{1}{R}\Big )\Big ]^{\f{1}{\gamma-1}}
\end{equation}
with similar corresponding expressions for $p_3(r)$
and $T_3(r)$.

Let us now focus our attention in the magnetic layer $r_b < r <r_t$.  The
hydrostatic equation there is
\begin{equation}
\label{eq:hydro2}
\f{d}{d{r}}\left[ p_2(r) + \frac{B^2(r)}{8 \pi} \right]  = -\f{\rho_2(r) 
GM_{\odot}}{r^{2}}.
\end{equation} 
Actually, there should also be a magnetic tension term which would break the spherical
symmetry.  Apart from breaking the spherical symmetry, the other problem which such a
term introduces is that it cannot be balanced by a combination of gradient forces, like
the pressure gradient and gravity (which comes from the gradient of gravitational potential).
For flux tubes at the bottom of SCZ, the magnetic tension term is supposed to be negligible
compared to other term (the gradient of magnetic pressure becomes much larger near the
boundaries of flux tubes).  Hence we have neglected this troublesome tension term.  We
further assume that the pressure and density inside the magnetic layer satisfy an
adiabatic relation  
\begin{equation}
\label{eq:state2}
p_{2}(r)=\kappa_{2} \rho_{2}(r)^{\gamma} 
\end{equation}
with a constant $\kappa_2$ which can be different from $\kappa$.  Since we do not expect
vigorous convection inside the magnetized region, the adiabatic assumption there is
somewhat questionable even though it simplifies our life.

To find the structure of the magnetic layer, we need to solve (9) and (10) subject
to certain conditions.  Since we want the pressure to be continuous across the
top and bottom of the magnetic layer, we should have 
\begin{equation}
\label{eq:press1}
p_1(r_{b})-p_{2}(r_{b})=\f{B^{2}(r_b)}{8 \pi},  
\end{equation}
\begin{equation}
\label{eq:press}
p_3(r_{t})-p_{2}(r_{t})=\f{B^{2} (r_t)}{8 \pi}. 
\end{equation}
Additionally, total mass of the system does not change while the magnetic
field is being created, implying
\begin{eqnarray}
\nonumber
\int_{R_{BCZ}}^{R_{\odot}} \rho_0(r) r^{2} dr & = &
\int_{R_{BCZ}}^{r_{b}} \rl (r) r^{2} dr + \int_{r_{b}}^{r_{t}} 
\rho_{2}(r) r^2 dr + \int_{r_{t}}^{R} \rp(r) r^2 dr \\ 
\label{eq:massconv1}
&=& \f{0.02 M_{\odot}}{4 \pi},
\end{eqnarray}
where we have assumed that only a total $0.02 M_{\odot}$ amount of mass is contained in
the outer regions of the Sun we are considering, justifying our approximating the
gravitational field by an inverse square law in (5).
It easily follows from (7) and (13) that
\begin{equation}
\kappa=\f{(\gamma-1)GM_{\odot}}{\gamma}\Big (\f{4\pi}{0.02M_{\odot}}\Big )
^{\gamma-1} R_{\odot}^{3\gamma-4} \Big(\int_{1/a}^{1} \xi^{\f{2\gamma-3}{\gamma-1}} 
(1-\xi)^{\f{1}{\gamma-1}} d\xi\Big)^{\gamma-1}
\end{equation} 
where $a= R_{\odot}/{R_{BCZ}}$ and $\xi$ is a dummy  non-dimensional 
variable. We would also like to specify a thermal condition of the
magnetic layer, i.e.\ to specify that its temperature is not very different from the
temperatures of the surroundings.  However, for the cases we considered, we
do not seem to have this freedom.

To solve the problem, we need to specify the magnetic field in the flux
layer.  The simplest possibility is to assume that we have a 
magnetic field of constant $B$ inside this layer.  Then (9) reduces to
the form (5) and can easily be integrated in combination with (10) to give
\begin{equation}
\rho_{2}(r)=\Big [\f{(\gamma-1)GM_{\odot}}{\gamma \kappa_{2}}
\Big (\f{1}{r}-\f{1}{R_{2}}\Big )\Big ]^{\f{1}{\gamma-1}},
\end{equation}
where $1/R_{2}$ is essentially the integration constant.  The expressions
for pressure and temperature can easily be written down.  We now have to 
satisfy the
conditions (11), (12) and (13).  If $B$ is assumed to be given, then $R$, $R_2$
and $\kappa_2$ have to be found such these three conditions are satisfied.  This
is a mathematically well-posed problem, since we have to find three quantities
from three equations.  It is straightforward to solve this problem numerically.
It may be noted that we can use the incomplete $\beta$ function defined as follows
\begin{equation}
\nonumber
\beta(x,\parc,\pard)= \f{1}{\beta(\parc,\pard)}\int_{0}^{x} \xi^{\f{2\gamma-3}{\gamma-1}}
(1-\xi)^{\f{1}{\gamma-1}} d\xi
\end{equation}
to write down (13) in the following form
\begin{eqnarray}
\label{eq:final}
\nonumber
{\!}\kappa^{-\f{1}{\gamma-1}}
\Rs^{\parc}\Big \{1-\beta(\f{r_{o}}{\Rs}
,\parc,\pard) \Big \}{\qquad}\\=\kappa_{2}^{-\f{1}{\gamma-1}}R_{2}^
{\f{3\gamma-4}{\gamma-1}}
\Big \{\beta(\f{r_{m}}{R_{2}},\f{3\gamma-4}{\gamma-1},
\f{\gamma}{\gamma-1})-\beta(\f{r_{o}}{R_{2}},\parc,\pard)\Big \}& &\\ \nonumber
+\kappa^{-\f{1}{\gamma-1}}
R^{\f{3\gamma-4}{\gamma-1}}\Big \{1-\beta(\f{r_{m}}{R},\parc,\pard)\Big \},
\end{eqnarray}
which facilitates the numerical computation. On solving the problem
numerically, we found that the radius increase was even much smaller than
what we found in our order-of-magnitude estimate, and what is more puzzling
is that the radius variation is less for a stronger magnetic field---a
clearly unphysical result! Figure~1a shows $\Delta R = R - R_{\odot}$ for
various values of $B$ (the magnetic flux has been kept the same, so a
larger $B$ implies a thinner magnetic layer). This unphysical result 
obviously is due to the
fact that the formulation used by us left us no freedom to ensure that the
temperature of the magnetic layer remained close to the temperature of the
surroundings.  In fact, $\kappa_2$ was found to be smaller than
$\kappa$, making the density in some regions of the magnetic layer more 
than the density in the surrounding region as shown in Figure~1b (with 
temperatures reduced inside the magnetic layer).  It is clear that Figure~1a
cannot be taken to present very realistic values of radius variation
with magnetic activity.
\begin{figure}
\centering
\includegraphics[height=7cm ,width=12cm]{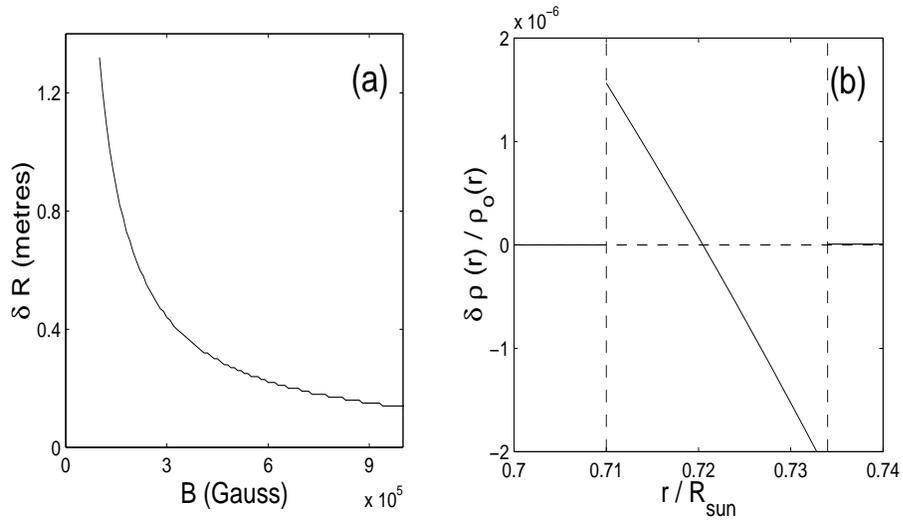}
\center{\caption{(a) The radius change $\delta R
= R-\Rs$ for various values of the magnetic field at the bottom of the convection zone. 
Counter-intuitively, $\delta R$ decreases with an increase in B.  (b) The 
fractional change  in density 
$\delta \rho(r)/{\rho_{o}(r)}$ inside the magnetic flux shell with $B= 10^{5}$ G (where 
$\rho_o(r)$ is the density in the absence of magnetic fields as given by (7)).
Note that for a region above $r_{b}$, the density 
is actually more than that below the shell!}}
\end{figure}

We next try to construct a solution by assuming the
plasma beta $\beta = 8 \pi p_{2}/B^{2}$ to be constant inside 
the magnetic shell. The hydrostatic equation (9) inside the magnetic shell now
takes the form
\begin{equation}
\label{eq:hydro2}
\f{\pa{p_{2}}}{\pa{r}} = -\f{GM_{\odot}\rho_{2}}{(1+\beta^{-1})r^{2}}
\end{equation}
 Consequently $\rho_{2}(r)$ has the expression
\begin{equation}
\label{eq:rho2}
\rho_{2}(r)= \Big [\f{(\gamma-1)GM_{\odot}}{\gamma \kappa_{2}(1+\beta^{-1})}
\Big (\f{1}{r}-\f{1}{R_{2}}\Big )\Big ]^{\f{1}{\gamma-1}}
\end{equation}
When we demanded that the conditions (11), (12) and (13) are satisfied,
we were able to find only one solution of the problem, which corresponds
to $\beta^{-1} = 0$ (i.e.\ no magnetic field) for which there is no 
change in radius.  This is certainly a consistent solution, but not
the solution we have been looking for.

Although we have not succeeded in constructing a satisfactory hydrostatic
model, we decided to present this brief discussion as a warning to future
researchers that the simplest and most obvious things do not work here. There
is no doubt that (9), (11), (12) and (13) should hold.  Perhaps (10) should
be replaced by a more realistic thermodynamic equation and we should use
the physics of magnetic field creation in more detail to decide how $B(r)$
may vary within the magnetic layer.  For these more complicated cases, we
shall not be able to write down an analytical solution of density as we
do in (15) or (18), and consequently the solution of the problem will be
much more complicated.  We felt that it will be pre-mature to attempt such
a solution now.  Given the uncertainties in the observational data, the
order-of-magnitude estimate should more than serve our purpose at present.

\section{Conclusion} 

Within the currently understood framework of solar MHD, we have not been
able to provide a theoretical explanation of the controversial claim 
made by Dziembowski {\em et al.}\ (2001) that the Sun shrinks with increasing
magnetic activity at the rate of 1.5 km yr$^{-1}$.  Spruit (1982, 1994)
estimated the radius change due to heat blocking by sunspots near the
solar surface.  Not only the radius variation was found to be much smaller,
the radius is expected to increase with the solar maximum.  We considered
the possibility if the excess magnetic pressure of flux tubes near the
base of the SCZ can give us a different result.  Even though traditional
dynamo models regard the solar maximum as the time when the magnetic flux
in the solar interior peaks and would suggest a radius increase at the
time of the solar maximum, we pointed out the model of Nandy and Choudhuri
(2002) provides a different scenario.  Although this model would predict
a decrease of the solar radius with increasing magnetic activity (i.e.\
with increasing loss of magnetic flux from the storage region), various
reasonable assumptions give a radius decrease rate about one order of
magnitude smaller than what is claimed by Dziembowski {\em et al}.\ (2001).  It
is true that our estimate is based on a very rough calculation.  However,
we believe that it is still an over-estimate rather than an under-estimate.
Only if the magnetic field at the base of the SCZ was concentrated to
values as high as $10^6$ G, the radius shrinkage would have
been as large as reported by Dziembowski {\em et al.}\ (2001).  There is no
theoretical reason to expect a magnetic field as strong as $10^6$ G
at the base of the SCZ.  In fact, it is not easy to generate
even a $10^5$ G magnetic field there (Choudhuri, 2003).  We conclude that
either the claim of Dziembowski {\em et al.}\ (2001) is incorrect, or else we do
not understand some basic physics of the solar cycle.
\section*{Acknowledgements}
We wish to thank S.\ C.\ Tripathy for valuable discussions.

%\end{article}
\end{document}